\documentstyle[12pt]{article}
\begin{document}
\renewcommand{\thesection}{\Roman{section}}
\baselineskip 20pt
\input feynman.tex
{\hfill PUTP-96-28}
\vskip 3cm
\begin{center}
{\large\bf A Crucial Test for Color-octet Production Mechanism in $Z^0$ Decays}
\end{center}
\vskip 7mm
\centerline{Cong-Feng Qiao,~~~Feng Yuan,~~~and~ Kuang-Ta Chao}
\centerline{\small\it Center of Theoretical Physics, CCAST(World Laboratory), 
            Beijing 100080, P.R. China}
\centerline{\small\it Department of Physics, Peking University, 
            Beijing 100871, P.R. China}
\begin{center}
\begin{minipage}{130mm}
\vskip0.6in
\begin{center}{\bf Abstract}\end{center}
  {The direct production rates of $D$-wave charmonia 
  in the decays of $Z^0$ is
  evaluated. The color-octet production processes 
  $Z^0\rightarrow ^3\!D_J(c\bar c) q\bar q$ are shown to have distinctively 
  large branching ratios, the same order of magnitude as that of $J/\psi$ 
  prodution, as compared with other $D$-wave charmonium 
  production mechanisms. This may suggest a crucial channel to test the
  color-octet mechanism as well as to observe the $D$-wave 
  charmonium states in
  $Z^0$ decays. In addition, a signal for the $^3D_J$ charmonium as strong as
  $J/\psi$ or $\psi^\prime$ with large transverse momentum at the Tevatron should
  also be observed.}
\vskip 1cm
PACS number(s):$12.38.Bx,$~ $14.40.Gx$
\end{minipage}
\end{center}

\vfill\eject\pagestyle{plain}\setcounter{page}{1}
The systematic study of the heavy quark bound systems has played a very 
important role in obtaining information not only on the properties of 
heavy quarks themselves but also on Quantum Chromodynamics(QCD). The 
property of asymptotic freedom  of QCD allows one to calculate the production 
and decays of heavy quark mesons perturbatively at high energy scale, while
the nonperturbative part can be factored out as the wave functions
or their derivatives at the origin.

Recent progress in the this area was stimulated by the experiment results of 
{\bf CDF} detactor at the Fermilab Tevatron. In the 1992-1993 run, the 
{\bf CDF} data\cite{1p} for the prompt production of $\psi$ and $\psi^\prime$
at large transverse momentum region were observed to be orders of magnitude
stronger than the lowest order perturbative calculations based on 
color-singlet model\cite{2p}, which has ever gained some success in describing the 
production and decays of heavy quarkonia.

To resolve these discrepancies, Braaten and Yuan suggested\cite{3p} that parton
fragmentation represents the dominant source of prompt quarkonium production 
at high 
transverse monentum ($P_T\ge 6~GeV$\cite{4p}), though these processes are
formally of high order in the strong coupling constant. Over the past few years,
the fragmentation functions for S- and P-wave quarkonia 
have been calculated to
lowest order\cite{3p}\cite{5p}\cite{6p}\cite{7p}\cite{8p}\cite{11p}, 
and they have subsequently been utilized to 
the issues of quarkonium production phenomenology in hadron collider. However,
they still underestimate the rates of $\psi$ and $\psi^\prime$ production
more than an order of magnitude referring to the observed data at the Tevatron\cite{4p}\cite{mat}.
These large discrepancies have called in question the simple color-singlet 
model description for quarkonium and stimulate ones to seek for new production
mechanisms as well as new paradigms for treating heavy quark-antiquark bound
systems that go beyond the color-singlet model.

To this end, a factorization formalism has recently been performed by Bodwin,
Braaten, and Lepage\cite{9p} in the context of nonrelativistic quantum 
chromodynamics(NRQCD), which provides a new framework to calculate the 
inclusive production and decay rates of quarkonia. In this approach, the 
calculations are organized 
in powers of $v$, the average velocity of the heavy quark(antiquark) 
in the meson rest frame, and in $\alpha_s$, the strong coupling
constant.

The breakdown of color-singlet model stems from its overlook of the high Fock
components contributions to quarkonium production cross sections. The 
color-octet term in the gluon fragmenting to $\psi(\psi^\prime)$ has been 
considered by Braaten and Fleming\cite{10p} to explain the 
$\psi(\psi^\prime)$ surplus problem discovered by ${\bf CDF}$.
Taking
$<{\cal O}^\psi_8(^3S_1)>$ and $<{\cal O}^{\psi^\prime}_8(^3S_1)>$ as input
parameters, the {\bf CDF} surplus problems for 
$\psi$ and $\psi^\prime$ can be explained as the contributions of color-octet 
terms due to gluon fragmentation. By adjusting the values for matrix elements
$<{\cal O}^\psi_8(^3S_1)>$ and $<{\cal O}^{\psi^\prime}_8(^3S_1)>$, the 
authors of Refs.\cite{10p}\cite{cho} got self-consistent values of them 
in the contents of NRQCD.

Even though the color-octet mechanism has gained some successes in describing
the production and decays of heavy quark bound systems, especially 
in explaining
the transverse momentum spectrum of the quarkonium measured at the Tevatron\cite{annrev},
it still has a long way to go before finally setting its position and role
in heavy quarkonium physics. Acturally, by now the
color-octet matrix elements are extracted only by fitting the data. Moreover,
some recent studies\cite{x1}\cite{x2}\cite{x3} show that there are
some inconsistencies about the normalization of the color-octet matrix
elements in different processes. Therefore, the most urgent task among  
others needs to do is to confirm and 
identify the color-octet quarkonium signals.

While the first charmonium state, the $J/\psi$, has 
been found over twenty years, $D$-wave states,
given the limited experimented data, have received less attention. However, 
this situation may be changed in both experimental 
and theoretical investigations. 
Experimentally, there are hopes of observing charmonium $D$-wave
states in addition to $1^{--}(\psi(3770))$ in a high-statistic exclusive
charmonium production experiment\cite{12p} and $b\bar{b}$~ $D$-wave states in
$\Upsilon$ radiative decays\cite{13p}. 
Theoretically, the analysis based on NRQCD shows that in $D$-wave quarkonium 
production the color-octet components play an even more important roles than
in the $S$- and $P$-wave cases.

Recently, there is some clue for the $D$-wave
$2^{--}$ charmonium state in $E705$ $300$ GeV $\pi^\pm$- and proton- Li 
interaction experiment\cite{14p}. In this 
experiment there is an abnormal phenomenon
that in the $J/\psi\pi^+\pi^-$ mass spectrum, two peaks at 
$\psi(3686)$ mass
and at $3.836~GeV$ (given to be the $2^{--}$ state) are observed and they
have almost the same
height. Obviously, 
this situation is difficult to explain based upon the color-singlet 
model, because the production rate would be proportional 
to the squared second
derivative of the wave function at the origin for the $D$-wave state, 
which is
suppressed by $O(v^4)$, as compared to the squared wave function at the origin
for the $S$-wave state. However, it might be explained 
with the NRQCD analysis. In NRQCD the
Fock state expansion for $^3D_J$ states is 
\begin{eqnarray}
\label{zankai}
|^3D_J>=O(1)|Q\bar{Q}(^3D_{J},\b{1})> + O(v)|Q\bar{Q}(^3P_{J'},\b{8})g>+
    O(v^2)|Q\bar{Q}(^3S_1,\b{8}~ or~ \b{1})gg>+\cdots.
\end{eqnarray}
In fact, considering the suppression due to the derivative of wave function
at the origin, in production processes, contributions of the 
three terms written out in Eq.(\ref{zankai}) 
have the same order in $v^2$. So in the quark
fragmentation, all of them are of the same order in both $\alpha_s$ and 
$v^2$. However, in the gluon fragmentation for $D$-wave charmonium production 
processes, the $S$-wave color-octet $(^3S_1,\b{8})$
production is $O(1/\alpha_s^2)$ enhanced over the color-singlet
$(^3D_J,\b{1})$ and $(^3S_1,\b{1})$ production in
the short distance perturbative sector
because the color singlet $(^3D_J,\b{1})$ and $(^3S_1,\b{1})$ 
have to couple to at
least three gluons. The $P$-wave color-octet process is forbidden
by charge conjugation invarance. By this argument, it might be easy 
to understand
the $E705$ experiment data as long as the nonperturbative matrix element
$<{\cal O}^{^3D_2}_8(^3S_1)>$ is about
the same order as $<{\cal O}^{\psi'}_8(^3S_1)>$ in magnitude, which is just 
expected in 
the framework of NRQCD.

Of course, this explanation requires the dominance of color-octet gluon
fragmentation over other production mechanisms. At energies in fixed
target experiments like $E705$, the color-octet gluon fragmentation dominance
may or may not be the case. Moreover, the strong signal of $J/\psi\pi^+\pi^-$
at $3.836 GeV$ observed by $E705$ is now questioned by other experiments\cite{e627}. Nevertheless, if the $E705$ result is confirmed 
(even with a smaller
rate, say, by a factor of 3, for the signal at $3.836 GeV$), the color-octet
gluon fragmentation will perhaps provide a quite unique explanation for the
$D$-wave charmonium production, since in all other mechanisms the $D$-wave
production rate is expected to be much smaller than that for the $S$-wave
states.

No matter what will be the final situation for the $E705$ result, 
it does remind
us that in the NRQCD approach the production rate of $D$-wave 
heavy quarkonium
states can be as large as that of $S$-wave states as long as the color-octet
gluon production mechanism dominantes. This scenario can be tested in many
processes, in particular, in the $Z^0$ decays.

Well over $10^6~Z^0$ events have been 
accumulated at the CERN $e^+e^-$ collider
at {\bf LEP} and further improvement are expected. This makes it possible to
investigate rare decays of $Z^0$ and to precisely test QCD.
Among others the production of charmonium states, in particular the $^3D_J$
states, in the $Z^0$ decays will be very interesting in testing 
the color-octet production mechanism.

Recently a study shows\cite{15p}
that the leading order color-octet process in $\alpha_s$, say 
$Z^0\rightarrow \psi g$, has a relatively 
small branching ratio because of the 
large momentum transfer, and this is also the case for the $D$-wave charmonium
production. The dominant color-octet processes for 
$Z^0\rightarrow ^3\!\!D_J q\bar q$ as well 
as $Z^0\rightarrow \psi q\bar q$ begin at order $\alpha_s^2$ as shown
in Fig.1. Here $q$ represents $u,~d,~s,~c~or~b$ qurks. From Ref.\cite{15p}
we readily have 
\begin{eqnarray}
\label{a10}
\nonumber
\Gamma(Z \rightarrow ^3\!\! D_{J} q\bar q ) & = &\Gamma(Z\rightarrow q\bar q)
\frac{\alpha_s^2(2m_c)}{36}\frac{<O^{^3D_{J}}_8(^3S_1)>}
{m_c^3}\big\{ 5(1-\xi^2)
-2\xi \ln \xi\\ 
\nonumber
&+& \big[2Li_2(\frac{\xi }{1+\xi })
 - 2 Li_2(\frac{1}{1+\xi })\\
&-& 2\ln(1+\xi )\ln\xi  + 3\ln\xi  + \ln^2 \xi \big]
(1+\xi)^2 \big\}
\end{eqnarray}
in the limit $m_q=0$, 
where $Li_2(x)=-\int\limits_0^x dt ~{\rm ln}(1-t)/t$ is the Spence function.

From Eq.(\ref{a10}) we can get the branching ratios of 
$Z^0\rightarrow ^3\!D_J q\bar q$.
In the numerical calculation, we take\cite{16p}\cite{cho}
\begin{equation}
m_c=1.5GeV,~~ M_{^3D_J}\approx 2m_c,~~\alpha_s(2m_c)=0.253
\end{equation}
and
\begin{equation}
\label{od8}
<{\cal O}_8^{^3D_2}(^3S_1)>
\approx <{\cal O}_8^{\psi^\prime}(^3S_1)>=4.6\times 10^{-3} GeV^3.
\end{equation}
Here, in NRQCD $<{\cal O}_8^{^3D_2}(^3S_1)>$ should be of the same order as$~$
$<{\cal O}_8^{J/\psi}(^3S_1)>$ or$~$ $<{\cal O}_8^{\psi^\prime}(^3S_1)>$, and
we just take (\ref{od8}) as a tentative value for it, where the value of
$<{\cal O}_8^{\psi^\prime}(^3S_1)>$ was determined by fitting the CDF data
for surplus production of $\psi^\prime$ at the Tevatron\cite{cho}.
From approximate heavy-quark spin symmetry relations, we have
$$
<{\cal O}_8^{^3D_1}(^3S_1)>\approx \frac{3}{5}
<{\cal O}_8^{^3D_2}(^3S_1)>,$$
\begin{equation}
<{\cal O}_8^{^3D_3}(^3S_1)>\approx \frac{7}{5}
<{\cal O}_8^{^3D_2}(^3S_1)>.
\end{equation}
Summing over all the quark flavors($q=u,~d,~s,~c,~b$), one may 
obtain the deacy widths 
\begin{eqnarray}
\sum\limits_q \Gamma(Z^0\rightarrow ^3\!\!D_1 q\bar q)
\approx 0.7\times 10^{-4} GeV,\nonumber\\ 
\sum\limits_q \Gamma(Z^0\rightarrow ^3\!\!D_2 q\bar q)
\approx 1.2\times 10^{-4} GeV,\nonumber\\ 
\sum\limits_q \Gamma(Z^0\rightarrow ^3\!\!D_3 q\bar q)
\approx 1.7\times 10^{-4} GeV, 
\end{eqnarray}
and the fraction ratios
\begin{eqnarray}
\frac{\Gamma(Z^0\rightarrow ^3\!\!D_1 q\bar q)}{\Gamma(Z^0\rightarrow q\bar q)}
=2.0\times 10^{-4},\nonumber\\
\frac{\Gamma(Z^0\rightarrow ^3\!\!D_2 q\bar q)}{\Gamma(Z^0\rightarrow q\bar q)}
=3.4\times 10^{-4},\nonumber\\
\frac{\Gamma(Z^0\rightarrow ^3\!\!D_3 q\bar q)}{\Gamma(Z^0\rightarrow q\bar q)}
=4.8\times 10^{-4}.
\end{eqnarray}

The dominant color-singlet processes occur as shown in Fig.2 and Fig.3. 
Corresponding to the quark fragmentation in Fig.2, the branching
ratios of $^3D_J$ production 
in color-singlet processes are $2.3\times 10^{-6}$,
$3.6\times 10^{-6}$, amd $1.7\times 10^{-6}$ for $J=1,~2,~3$, respectively, 
which are obtained from the universal fragmentation calculations\cite{xxx}.
There should also be color-octet processes through quark 
fragmentation as in Fig.2.
However, indirect evidence indicates that they are 
not donimant relative
to the color-singlet processes\cite{xxx}.

The processes in Fig.3 are more complicated.
For in the most important kinematic region 
the virtual gluon is nearly on its massshell,
$^3D_J$ production in the gluon fragmentation 
color-singlet process may be separated to be $Z^0
\rightarrow q\bar q g^*$ with $g^*\rightarrow ^3\!\!D_J gg$. We can estimate 
the partial width following the way in Ref.\cite{18p}, and the differencial
decay rate of $Z^0\rightarrow q\bar q g^*$ may then be obtained. With
the definition 
\begin{equation}
  \Gamma(g^*\rightarrow AX)=\pi\mu^3 P(g^*\rightarrow AX),
\end{equation}
the calculation of decay distribution 
$P(g^*\rightarrow ^3\!\!D_Jgg)$ for the gluon of virtuality
$\mu$ is very complicated and lengthy (the detailed calculation will be
given elsewhere\cite{qyc}), and
in the nonrelativistic limit it is proportional to the 
second derivative of the radial 
wave function at the origin. As in the cases of  
$P$-wave charmonium 
production, $g^*\rightarrow ^3\!\!D_J gg$ 
processes also have the infrared divergences
involved, which are associated with the soft gluon in the final state. Strictly
speaking, the divergences can be cancelled in the framework of NRQCD, but 
here we simply deal with it following the way of\cite{7p} by 
imposing a lower cutoff
$\Lambda$ on the energy of the outgoing 
gluon in the quarkonium rest frame. As discussed
in Ref.\cite{7p}, the cutoff $\Lambda$ can be set to be $m_c$ to avoid large
logarithms in the divergent terms.

The decay widths of $Z^0$ to color-singlet charmonium state $^3D_J$ 
by gluon fragmentation can be evaluated via
\begin{equation}
\Gamma(Z^0\rightarrow q\bar q g^*;g^*\rightarrow ^3\!D_J gg)=
  \int \limits_{\mu_{min}^2}^{M_Z^2} d\mu^2 \Gamma(Z^0\rightarrow q\bar q g^*)
    P(g^*\rightarrow^3\!D_J gg),
\end{equation}
where the cutoff $\Lambda=m_c$ is transformed into a lower limit on $\mu^2_{min}=
12 m_c^2$.

In the numerical calculation, taking\cite{16p}\cite{20p}
\begin{equation}
\alpha_s(2m_c)=0.253,~~m_c=1.5 GeV,~~|R_D^{\prime\prime}(0)|^2=0.015 GeV^7,
\end{equation}
and summing over all the flavors $q~(q=u,d,s,c,b)$, we obtain
\begin{eqnarray}
\frac{\Gamma(Z^0\rightarrow q\bar q g^*;g^*\rightarrow ^3\!\!D_1 X)}
 {\Gamma(Z^0\rightarrow q \bar q)}=4.3\times 10^{-7},\nonumber\\
\frac{\Gamma(Z^0\rightarrow q\bar q g^*;g^*\rightarrow ^3\!\!D_2 X)}
 {\Gamma(Z^0\rightarrow q \bar q)}=2.1\times 10^{-6},\nonumber\\
\frac{\Gamma(Z^0\rightarrow q\bar q g^*;g^*\rightarrow ^3\!\!D_3 X)}
 {\Gamma(Z^0\rightarrow q \bar q)}=1.2\times 10^{-6}.
\end{eqnarray}

Among the three triplet states of $D$-wave charmonium, $^3D_2$ is the 
most prominant candidate
to discover firstly. Its mass falls in the range of $3.810\sim 3.840~GeV$ in the 
potential model calculation\cite{l1}\cite{l2}, that is above the
$D{\bar D}$ threshold but below the $D{\bar D}^*$
threshold. However the parity conservation forbids it decaying into $D\bar D$. 
It, therefore, is a narrow resonance. Its main decay modes are expected to be,
\begin{equation}
^3D_2\rightarrow J/\psi\pi\pi,~~~^3D_2\rightarrow ^3P_J \gamma(J=1,2),~~~
^3D_2\rightarrow 3g.
\end{equation}
We can estimate the hadronic transition 
rate of $^3D_2\rightarrow J/\psi\pi^+\pi^-$
from the Mark III data for 
$\psi(3770)\rightarrow J/\psi\pi^+\pi^-$\cite{zhu} and
the QCD multipole expansion theory\cite{yan}\cite{ky}. The Mark III data
give\cite{zhu}
$\Gamma(\psi(3770)\rightarrow J/\psi\pi^+\pi^-)=(37\pm 17\pm 8)~ KeV~~~
{\rm or}~~~(55\pm 23\pm 11)~ KeV$
(see also Ref.\cite{ky}). Because the $S-D$ mixing angle for $\psi(3770)$ and
$\psi(3686)$ is expected to be small 
(say, $-10^\circ$, see Ref.\cite{ding} for
the reasoning), 
the observed $\psi(3770)\rightarrow J/\psi\pi^+\pi^-$ transition
should dominantly come from the $^3D_1\rightarrow J/\psi\pi^+\pi^-$ transition,
which is also compatible with the multipole expansion estimate\cite{ky}.
Then using the relation\cite{yan}
$$
d\Gamma(^3D_2\rightarrow ^3S_1 2\pi)=d\Gamma(^3D_1\rightarrow ^3S_1 2\pi)
$$
and taking the average value of 
the $\Gamma(\psi(3770)\rightarrow J/\psi\pi^+\pi^-)$
from the Mark III data, we may have
\begin{equation}
\label{y1}
\Gamma(^3D_2\rightarrow J/\psi \pi^+\pi^-)=\Gamma(^3D_1\rightarrow J/\psi \pi^+\pi^-)
\approx 46~ KeV.
\end{equation}
For the E1 
transition $3D_2\rightarrow ^3P_J\gamma(J=1,2)$, using the potential
model with relativistic effects being considered\cite{dchao}, we find
\begin{equation}
\label{y2}
\Gamma(^3D_2 \rightarrow \chi_{c1}\gamma)=250~ KeV,~~~
\Gamma(^3D_2 \rightarrow \chi_{c2}\gamma)=60~ KeV,
\end{equation}
where the mass of $^3D_2$ is set to be $3.84 GeV$. 
As for the $^3D_2\rightarrow
3g$ annihilation decay, an estimate gives\cite{be}
\begin{equation}
\label{y3}
\Gamma(^3D_2\rightarrow 3g)=12~ KeV
\end{equation}
From (\ref{y1}), (\ref{y2}), and (\ref{y3}), we find
\begin{eqnarray}
\nonumber
\Gamma_{tot}(^3D_2)&\approx & \Gamma(^3D_2\rightarrow J/\psi \pi\pi)
+\Gamma(^3D_2 \rightarrow \chi_{c1}\gamma)+\Gamma(^3D_2 \rightarrow \chi_{c2}\gamma)
+\Gamma(^3D_2\rightarrow 3g)\\
&\approx &390~ KeV,
\end{eqnarray}
and
\begin{equation}
B(^3D_2\rightarrow J/\psi \pi^+\pi^-)\approx 0.12.
\end{equation}
Compared with $B(\psi^\prime\rightarrow J/\psi \pi^+\pi-)=0.324\pm 0.026$,
the branching ratio of $^3D_2\rightarrow J/\psi \pi^+\pi^-$ is only smaller
by a factor of 3, and therefore the decay mode of $^3D_2\rightarrow J/\psi \pi^+\pi^-$
is observable, 
if the production rate of $^3D_2$ is of the same order as
$\psi^\prime$.

The other two states of the triplet,
$^3D_1$ and $^3D_3$, are above the open channels threshold and are not narrow,
and therefore are difficult to detect. It might be interesting to note that
the {\bf OPAL} Collaboration at {\bf LEP} has analysed the $J/\psi\pi^+\pi^-$
spectrum in $Z^0$ decays\cite{opal}, and there seems to be some events above
the background around $3.77 GeV$, whether these events are associated with the
$D$-wave $1^{--}$ charmonium state $\psi(3770)$ might remain interesting.

In conclusion, from the calculations and discussions above 
one can clearly see that 
the branching ratios of gluon fragmenting to 
color-octet $^3D_J$ states are two
to three orders larger than the dominant color-singlet processes.
This large divergences are much helpful in distinguishing the two production
mechanisms in experiment.  On the other hand, because the production 
rates of $\psi^\prime$ and $^3D_2(c\bar c)$ 
in color-octet mechanism
are of the same amount of magnitude, the $2^{--}$
charmonium production in $Z^0$ decay may 
provide a crucial channel to test color-octet production mechanism
at {\bf LEP} with present luminosity.
As for the $b\bar b$ system,, the production of $D$-wave bottomonium states by color-singlet
and color-octet mechanisms are similar to that of charmonium, but the 
disparity of the branching ratios between these two mechanisms is not as 
evident as that of charmonium production.

Finally, we would like to point out that the above discussion also applies to
the $^3D_J$ production at the Tevatron. As in the case of $J/\psi$ and $\psi^\prime$
production at large momentum transverse, the $^3D_J$ production will also
be dominanted by the color-octet gluon fragmentation. This
implies that a signal for $^3D_J$ states, which can be as strong as $J/\psi$
or $\psi^\prime$, should be observed at the Tevatron. This will also be a crucial
test of the color-octet mechanism. Detailed analysis will be given elsewhere.

\vskip 1cm
\begin{center}
\bf\large\bf{Acknowlegement}
\end{center}

This work was supported in part by the National Natural Science Foundation
of China, the State Education Commission of China and the State Commission
of Science and Technology of China.

\newpage
\centerline{\bf \large Figure Captions}
\vskip 2cm
\noindent
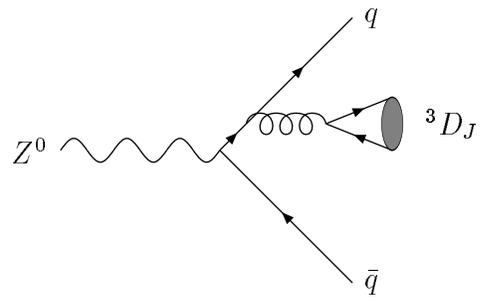
Fig.1. One of the contributing  Feynman diagrams of color-octet mechanism in 
$Z^0\rightarrow ^3\!\!D_J q\bar{q}$ processes.\\
\noindent 
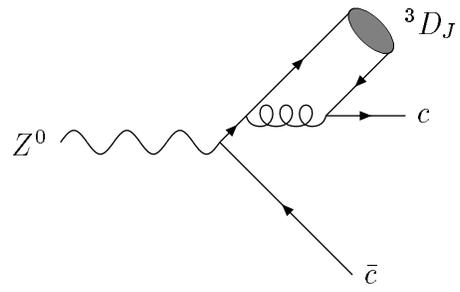
Fig.2. One of the Feynman diagrams corresponding to quark fragmentation   
processes in $Z^0$ decays.\\
\noindent 
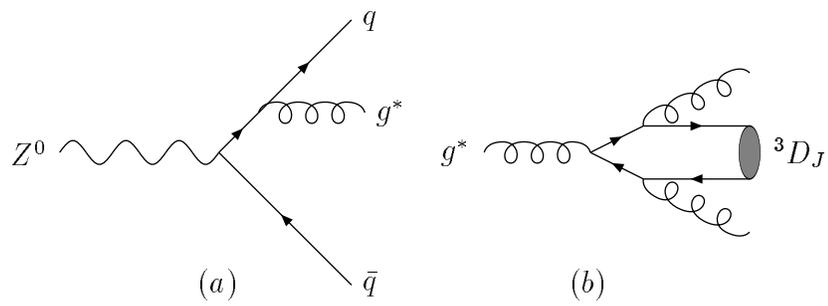
Fig.3. Diagrams for $^3D_J$ production from gluon jet in color-singlet 
mechanisms. (a)virtual gluon production in $Z^0$ decays (b)$^3D_J$ production  
in gluon fragmentaion.

\end{document}